\documentclass[aps, prb, twocolumn, amssymb, amsmath, showpacs, superscriptaddress]{revtex4-1}
\usepackage{bm}
\usepackage{times}
\usepackage{graphicx}
\usepackage{color}
\usepackage{dcolumn}
\usepackage[colorlinks=true, pdfstartview=FitV, linkcolor=blue, citecolor=blue, urlcolor=blue]{hyperref}
\usepackage{appendix}
\usepackage[normalem]{ulem}

\begin{document}
\title{Quantum Fluctuation of the Quantum Geometric Tensor and its Manifestation as Intrinsic Hall Signatures in Time-Reversal Invariant Systems}
\author{Miaomiao Wei$^\S$}
\affiliation{College of Physics and Optoelectronic Engineering, Shenzhen University, Shenzhen 518060, China}
\author{Luyang Wang$^\S$}
\affiliation{College of Physics and Optoelectronic Engineering, Shenzhen University, Shenzhen 518060, China}
\author{Bin Wang}
\affiliation{College of Physics and Optoelectronic Engineering, Shenzhen University, Shenzhen 518060, China}
\author{Longjun Xiang}
\affiliation{College of Physics and Optoelectronic Engineering, Shenzhen University, Shenzhen 518060, China}
\author{Fuming Xu}
\email[]{xufuming@szu.edu.cn}
\affiliation{College of Physics and Optoelectronic Engineering, Shenzhen University, Shenzhen 518060, China}
\author{Baigeng Wang}
\affiliation{National Laboratory of Solid State Microstructures and Department of Physics, Nanjing University, Nanjing 210093, China}
\affiliation{Collaborative Innovation Center for Advanced Microstructures, Nanjing 210093, China}
\author{Jian Wang}
\email[]{jianwang@hku.hk}
\affiliation{College of Physics and Optoelectronic Engineering, Shenzhen University, Shenzhen 518060, China}
\affiliation{Department of Physics, The University of Hong Kong, Pokfulam Road, Hong Kong, China}

\begin{abstract}
In time-reversal invariant systems, all charge Hall effects predicted so far are extrinsic effects due to the dependence on the relaxation time. We explore intrinsic Hall signatures by studying quantum noise spectrum of the Hall current in time-reversal invariant systems, and discover intrinsic thermal Hall noises in both linear and nonlinear regimes. As the band geometric characteristics, quantum geometric tensor and Berry curvature play critical roles in various Hall effects, so are their quantum fluctuations. It is found that the thermal Hall noise in linear order of the electric field is purely intrinsic, and the second-order thermal Hall noise has both intrinsic and extrinsic contributions. In particular, the intrinsic part of the second-order thermal Hall noise is a manifestation of the quantum fluctuation of quantum geometric tensor, which widely exists as long as Berry curvature is nonzero. These intrinsic thermal Hall noises provide direct measurable means to band geometric information, including Berry curvature related quantities and quantum fluctuation of quantum geometric tensor.
\end{abstract}
\maketitle

\noindent{\it Introduction.} Quantum geometric tensor (QGT)\cite{Provost,Polkovnikov,Bleu,Asteria,Gianfrate} has fundamental importance in modern condensed matter physics, since it contains the band geometry and topology information of the underlying Hamiltonian. The imaginary part of QGT, Berry curvature,\cite{Niu} plays critical role in various Hall effects, such as quantum Hall effect,\cite{Klitzing} quantum spin Hall effect,\cite{BHZ,Konig} and quantum anomalous Hall effect,\cite{Haldane,Nagaosa,Xue} etc. The dipole moment of Berry curvature, i.e., Berry curvature dipole (BCD),\cite{L-Fu,Guinea1} can induce an \textit{extrinsic} second-order nonlinear Hall effect in time-reversal (TR) invariant systems, which has been intensively discussed\cite{J-You,Du,Lee,Ortix,Sodemann,Tsymbal,AdvQuantum2021,WeiMM} and experimentally verified.\cite{S-Xu,Q-Ma,K-Kang,J-Xiao,NHEBi2Se3,H-Yang,NatElec2021} It is found that extrinsic mechanisms such as skew-scattering can dominate the second-order Hall effect in thick T$_d$-MoTe2 samples and result in giant $c$-axis nonlinear Hall conductivity.\cite{Tiwari21} Intrinsic response properties, which is independent of the relaxation time and not affected by the scattering process, can directly probe band information related to Berry curvature as well as QGT, and hence are of special interest. Up to now, intrinsic linear spin Hall effect has been proposed in both TR-invariant systems\cite{sinova} and TR-broken systems.\cite{spinHall2011} In nonlinear regime, intrinsic second-order anomalous Hall effects have been reported in TR-broken systems recently.\cite{niu2014,dxiao,syyang} As far as we know, in TR-invariant systems, intrinsic charge Hall transport phenomenon induced by Berry curvature or QGT has not been predicted.

The Hall effect refers to the transverse current or voltage in response to the driving electric field. In various Hall transport, quantum fluctuation of currents widely exists. Such quantum fluctuation or quantum noise originates from the quantum nature of charge carriers, where quantum interference and Pauli exclusion principle play important roles.\cite{but1} Therefore, in addition to the average current, quantum noise associated with the Hall current, i.e., the Hall/transverse noise spectrum, and higher-order correlations,\cite{Levitov,Tang} are needed to fully characterize Hall transport properties, as well as the underlying Berry curvature and QGT. In topological insulators, quantum noise has been proposed to assess the quality of edge state transport,\cite{Nagaev2016,Glazman2017,Goldstein2019,Burmistrov2020} which are experimentally measured in HgTe quantum wells\cite{Dvoretsky2015,Tikhonov2017} and InAS/Ga(In)Sb structures.\cite{Natelson2019} Quantum noise has also been utilized to identify strongly correlated nonlocal Majorana states.\cite{SCZhang2011,Oreg2015,Xie2018,Martin2019,Qiao2019,Simon2021} On the other hand, there are attempts to measure QGT and related topological matters,\cite{Mudry2013,Ozawa2018,Goldman2018} via superconducting qubit\cite{SLZhu2019} or microwave spectroscopy.\cite{Belzig2020}

Several questions arise: is there any intrinsic charge Hall signature in time-reversal invariant systems? how to describe quantum noise of the Hall current and fluctuation of the quantum geometric tensor? what is the relation between them?

In this work, we answer these questions by studying quantum current correlation in coherent transport and identifying its relation to QGT. In TR-invariant systems, when expanding the thermal noise spectrum in terms of the electric field at low voltage, we find intrinsic thermal Hall noises in both linear and nonlinear response regimes, which are explicitly expressed in terms of Berry curvature related quantities. The linear thermal Hall noise is purely intrinsic, and an interesting dual relation on BCD is established between the extrinsic second-order Hall current and this intrinsic linear Hall noise in two-dimensional (2D) systems. The intrinsic part of the second-order thermal Hall noise is contributed by quantum fluctuation of Berry curvature, which is the manifestation of quantum fluctuation of QGT. The Hall noise spectrum can be easily measured on platforms similar to previous detection of the BCD-induced second-order Hall effect,\cite{S-Xu,Q-Ma,K-Kang,J-Xiao,NHEBi2Se3,H-Yang} and we propose feasible strategies to extract Berry curvature fluctuation from the noise signals. Berry curvature fluctuation in 3D systems is also discussed, which is directly accessible via a three-step measurement. These findings reveal that, QGT related information, i.e., Berry curvature dipole and Berry curvature fluctuation, are qualitatively measurable through the intrinsic thermal Hall noises of TR-invariant systems.

\bigskip
\noindent{\it Hall noise spectrum.} Considering a system with TR symmetry $\cal T$ and nonzero Berry curvature, we examine the noise spectrum of the Hall current, i.e., the Hall noise spectrum. We start with the current density operator for dc transport ($\hbar=e=1$)
\begin{eqnarray}
{\hat J}_a = \sum_{mn} \int_k {\hat a}^\dagger_m {\hat a}_n (v^{nm}_a - \varepsilon_{abc} \Omega^{nm}_{b} E_c).
\end{eqnarray}
Here ${\hat a}^\dagger_m$ is the creation operator such that $\langle {\hat a}^\dagger_m {\hat a}_n \rangle = f_m \delta_{mn}$ with $f_m$ the nonequilibrium Fermi distribution function for band $m$. When $n = m$, $v^{nn}_a \equiv v^{n}_a = \partial \epsilon_n / \partial k_a$ is the group velocity with $\epsilon_n$ the band energy. When $n \neq m$, $v^{nm}_a = i(\epsilon_n-\epsilon_m)A^a_{nm}$ is the interband velocity matrix,\cite{niu2014} with $A^a_{nm}=\langle n|i\partial_{k_a}|m\rangle$ the interband Berry connection. $\varepsilon_{abc}$ is the Levi-Civita tensor. Similarly, Berry curvature $\Omega^{nm}_{c} \equiv \frac{1}{2} \varepsilon_{abc} \sum_l i(A^a_{nl} A^b_{lm} - A^b_{nl} A^a_{lm}) \bar{\delta}_{ln}\bar{\delta}_{lm}$ ($\bar{\delta}_{ln}=1-{\delta}_{ln}$) contains both intraband ($n = m$) and interband ($n \neq m$) contributions.\cite{QianXF2022} $a$, $b$, and $c$ label $x$, $y$, and $z$ in Cartesian coordinates. Thus $E_c$ stands for the electric field in $c$ direction.

Quantum correlation of the current density is defined as\cite{but1} ${\delta(0)} S_{ab}=\langle (\Delta {\hat J}_a )(\Delta {\hat J}_b) \rangle$, where $\Delta {\hat J}_a= {\hat J}_a - \langle {\hat J}_a \rangle$ and $\langle {\hat J}_a \rangle$ is the expectation value of $ {\hat J}_a$. The quantum noise has two contributions. One vanishes at zero temperature and is referred as thermal noise,
\begin{eqnarray}\label{eq0}
S^{T}_{ab}=\sum_n\int_k f_n (1 - f_n) (v_a^n - \varepsilon_{abc} \Omega_b^n E_c) \nonumber \\
           \times (v_b^n - \varepsilon_{bb_1 c_1} \Omega_{b_1}^n E_{c_1}),
\end{eqnarray}
The other corresponds to the shot noise,
\begin{eqnarray}
S^{S}_{ab} =\frac{1}{2} \sum_{m\ne n}\int_k \bar{f}_{mn} (v_a^{nm} - \varepsilon_{abc} \Omega_b^{nm} E_c) \nonumber \\
            \times (v_b^{mn} - \varepsilon_{bb_1 c_1} \Omega_{b_1}^{mn} E_{c_1}),
\end{eqnarray}
where the factor $\bar{f}_{mn}=f_m(1-f_n)+f_n(1-f_m)$ ensures that $S^{S}_{ab}$ is finite at zero temperature. The detection of shot noise requires low temperature, where $S^{S}_{ab}$ dominates the noise spectrum in the high voltage regime,\cite{but1} i.e., $eEl \gg k_B T$ with $l$ the system size. In contrast, thermal noise is dominant at low voltage $eEl \ll k_BT$\cite{but1} and easily measurable via temperature-dependent experiments in large temperature ranges, from several Kelvin\cite{Henny1999} to room temperature.\cite{Birk1995} In the following, we focus on the thermal noise in the regime $eEl \ll k_BT$. For simplicity, $S^{T}_{ab}$ is denoted as $S_{ab}$ and all noises discussed below are thermal noises.

For a particular band $n$,\cite{Fnote0} Eq.(\ref{eq0}) can be transformed into a compact vector form,
\begin{eqnarray}
S_{ab} = \int_k  f_n(1-f_n) ({\bf v} - {\bf \Omega} \times {\bf E})_a ({\bf v} - {\bf \Omega} \times {\bf E})_b. \label{eq3}
\end{eqnarray}
$S_{ab}$ is a second-rank symmetric tensor. Its diagonal elements, $S_{aa}$ ($a=x,y$, and $z$), are auto-correlation of currents and contain the Hall noises. The off-diagonal elements, $S_{ab} = S_{ba}$ ($a \neq b$), correspond to the cross-correlation function. Expanding $S_{ab}$ in terms of the electric field in the regime $eEl \ll k_B T$, we have
\begin{eqnarray}
S_{ab} = S^{(1)}_{ab}+ S^{(2)}_{ab} + {\cal O}(E^3)... ,
\end{eqnarray}
where $S^{(1)}_{ab}$ and $S^{(2)}_{ab}$ are the linear and second-order noise in electric field, respectively.

The linear noise is obtained from Eq.(\ref{eq3})
\begin{equation}
S^{(1)}_{ab} =  -k_B T \int_k  f_0 [\partial_a ({\bf \Omega} \times {\bf E})_b + \partial_b ({\bf \Omega} \times {\bf E})_a]. \label{s1}
\end{equation}
Here $\partial_a \equiv \partial_{k_a} $ and $f_0$ is the equilibrium distribution function. Denoting the BCD pseudotensor in matrix form as $D_{ab} = ({\bf D}_x,{\bf D}_y,{\bf D}_z)^T$ with ${\bf D}_a = \int_k f_0 \partial_a {\bf \Omega}$, we find
\begin{eqnarray}
S^{(1)}_{ab} = - k_B T ({\bf D}_a \times {\bf E})_b - k_B T ({\bf D}_b \times {\bf E})_a. \label{s1-3}
\end{eqnarray}
Symmetry analysis on $S^{(1)}_{ab}$ is presented in Sec.{\color{blue}I}(3) of the supplementary material. In 2D, point groups supporting nonzero $S^{(1)}_{ab}$ are $\{C_1,C_{1v},C_{2}\}$, while in 3D they are $\{C_n,C_{nv},D_{2},D_{2d},D_{3},D_{4},D_{6},S_{4} \}$ with $n=1,2,3,4,6$.

In 2D TR-invariant systems, only $\Omega_z$ can be nonzero and the highest symmetry allowed for nonvanishing BCD is single mirror symmetry.\cite{L-Fu} Consequently, BCD tensor $D_{ab}$ is reduced to an in-plane pseudovector. Labeling the BCD vector as ${\bf D} = \int_k f_0 \nabla_{\mathbf{k}} \Omega_z = (D_x, D_y)$, the diagonal element of $S^{(1)}_{ab}$, i.e. the linear Hall noise, is written as
\begin{eqnarray}
S^{(1)}_{aa} = 2k_B T D_a ~ {\hat z} \cdot ({\hat{a}} \times {\bf E}). \label{s1-2d}
\end{eqnarray}

For the same system, linear Hall effect vanishes due to the time-reversal constraint. Hence the leading order Hall effect is the BCD-induced second-order Hall current\cite{L-Fu}
\begin{eqnarray}
{\bf J}^{(2)} = \tau {\hat z} \times {\bf E} ({\bf D} \cdot {\bf E}). \label{NLcurrent}
\end{eqnarray}
In contrast to the {\it{extrinsic}} second-order Hall current which scales with the relaxation time $\tau$, the linear Hall noise $S^{(1)}_{aa}$ is an {\it{intrinsic}} property. The vector notation of linear Hall noise is highly relevant to that of the second-order Hall current. The nonlinear Hall current is optimal when electric field ${\bf E}$ is aligned with BCD vector ${\bf D}$, whereas vanishes for ${\bf E} \perp {\bf D}$. On the contrary, $S^{(1)}_{aa}$ is nonzero when the electric field is perpendicular to ${\bf D}$.

Using $g_2=f_0(1-f_0)=-k_B T \partial_\epsilon f_0$, the second-order noise is expressed as
\begin{eqnarray}
S^{(2)}_{ab}=\tau^2\int_k [\partial^2_{c d} g_2 + \partial_c f_0 \partial_d f_0] v_a v_b E_c E_d + k_B T {\cal E}_a^T \Omega^{(2)}_{ab} {\cal E}_b. \nonumber
\end{eqnarray}
Here $\mathcal{E}_a={\bf E} \times \hat{a}$ with $a=x,y$, and $z$. The second term of $S^{(2)}_{ab}$ is purely intrinsic and determined by $\Omega^{(2)}_{ab}$, a second-rank symmetric tensor defined as $\Omega^{(2)}_{ab} = \int_k (-\partial_\epsilon f_0) \Omega_a \Omega_b $. Notice that $\Omega^{(2)}_{ab}$ contains both auto-correlation of Berry curvature ($a = b$) as well as cross-correlation of Berry curvature ($a \ne b$) which is nonzero only in 3D. We term $\Omega^{(2)}_{ab}$ as the {\it Berry curvature fluctuation}, a new signature of band geometry, which is the second-order moment of Berry curvature. $\Omega^{(2)}_{ab}$ is observable in any systems with nonzero Berry curvature, and hence intrinsic contributions from $\Omega^{(2)}_{ab}$ to the second-order noise always exist in the presence or absence of $\cal T$ symmetry. We show in the following that $\Omega^{(2)}_{ab}$ is the manifestation of quantum fluctuation of QGT.

\bigskip
\noindent{\it Quantum fluctuation of QGT.} From Refs.[\onlinecite{Berry}] and [\onlinecite{Graf}], QGT operator for band $n$ is defined as
\begin{eqnarray}
{\hat T}_{nab} = \partial_a {\hat P}_n (1-{\hat P}_n)  \partial_b {\hat P}_n, \label{eq1}
\end{eqnarray}
where ${\hat P}_n = |n\rangle \langle n|$.\cite{Juan} We focus on 2D two-band systems and define QGT operator for the lower band as
\begin{equation}
{\hat T} = \left( \begin{array}{cc}
{\hat g}_{xx} & {\hat g}_{xy}-i(1/2){\hat \Omega}_{xy} \\
{\hat g}_{yx}-i(1/2){\hat \Omega}_{yx}  & {\hat g}_{yy}
\end{array} \right),
\end{equation}
which is a super-operator with components
\begin{eqnarray}
{\hat g}_{ab} &=& (1/2)(A^a_{01} A^b_{10}|0 \rangle \langle 0| + A^a_{10} A^b_{01}|1 \rangle \langle 1|), \label{g1} \\
{\hat \Omega}_{ab} &=& i(A^a_{01} A^b_{10}|0 \rangle \langle 0| - A^a_{10} A^b_{01}|1 \rangle \langle 1| ).\label{o1}
\end{eqnarray}

Clearly, the real part of QGT is the quantum metric $g_{ab}$, which characterizes the distance between quantum states.\cite{Ozawa2018,Graf} Here ${\hat T}$ and ${\hat g_{ab}}$ (denoted as ${\hat O}$) are Hermitian satisfying ${\hat O}_{ii}^\dagger = {\hat O}_{ii}$ and ${\hat O}_{ij}^\dagger = {\hat O}_{ji}$, while ${\hat \Omega_{ab}}$ is anti-Hermitian. Defining quantum fluctuation $\Delta O = \langle {\hat O}^2\rangle - \langle {\hat O}\rangle^2$, we obtain (Sec.{\color{blue}III} of the supplementary material\cite{sup})
\begin{eqnarray}
(\Delta g)_{xx} &=& (1/2) g_{xx}^2 -(\Delta \Omega)_{xx}, \nonumber\\
(\Delta g)_{yy} &=& (1/2) g_{yy}^2 -(\Delta \Omega)_{yy}, \nonumber\\
(\Delta g)_{xy} &=& (1/2) (g_{xx} + g_{yy}) g_{xy}, \nonumber
\end{eqnarray}
and
\begin{eqnarray}
(\Delta \Omega)_{xx} &=& (\Delta \Omega)_{yy}= -(1/2) (g_{xy}^2 + \Omega_{xy}^2). \nonumber
\end{eqnarray}
Here the minus sign reflects the anti-Hermitian nature of ${\hat \Omega}_{ab}$. Then we integrate $(\Delta \Omega)_{xx}$ and $(\Delta \Omega)_{yy}$ over the Brillion zone to obtain the observable fluctuation: $\Delta \Omega = \int_k f_0(\epsilon)[(\Delta \Omega)_{xx} + (\Delta \Omega)_{yy})]$. The quantum fluctuation of Berry curvature has precise correspondence with the semiclassical result: $\partial_\epsilon (\Delta \Omega) = \int_k (-\partial_\epsilon f_0) \Omega^{2}_{xy} = \int_k (-\partial_\epsilon f_0) \Omega^{2}_{z} = \Omega^{(2)}_{z}$, when $g_{xy} = 0$ for all $k$ points. Two notations $\Omega_{xy}$ and $\Omega_{z}$ are interchangeable.\cite{Niu} We provide several ways in Sec.{\color{blue}III} of the supplementary material\cite{sup} to design such Hamiltonian fulfilling $g_{xy} = 0$. Nevertheless, quantum fluctuation of QGT is manifested by quantum fluctuation of Berry curvature.

\begin{figure}[tbp]
\centering
\includegraphics[width=\columnwidth]{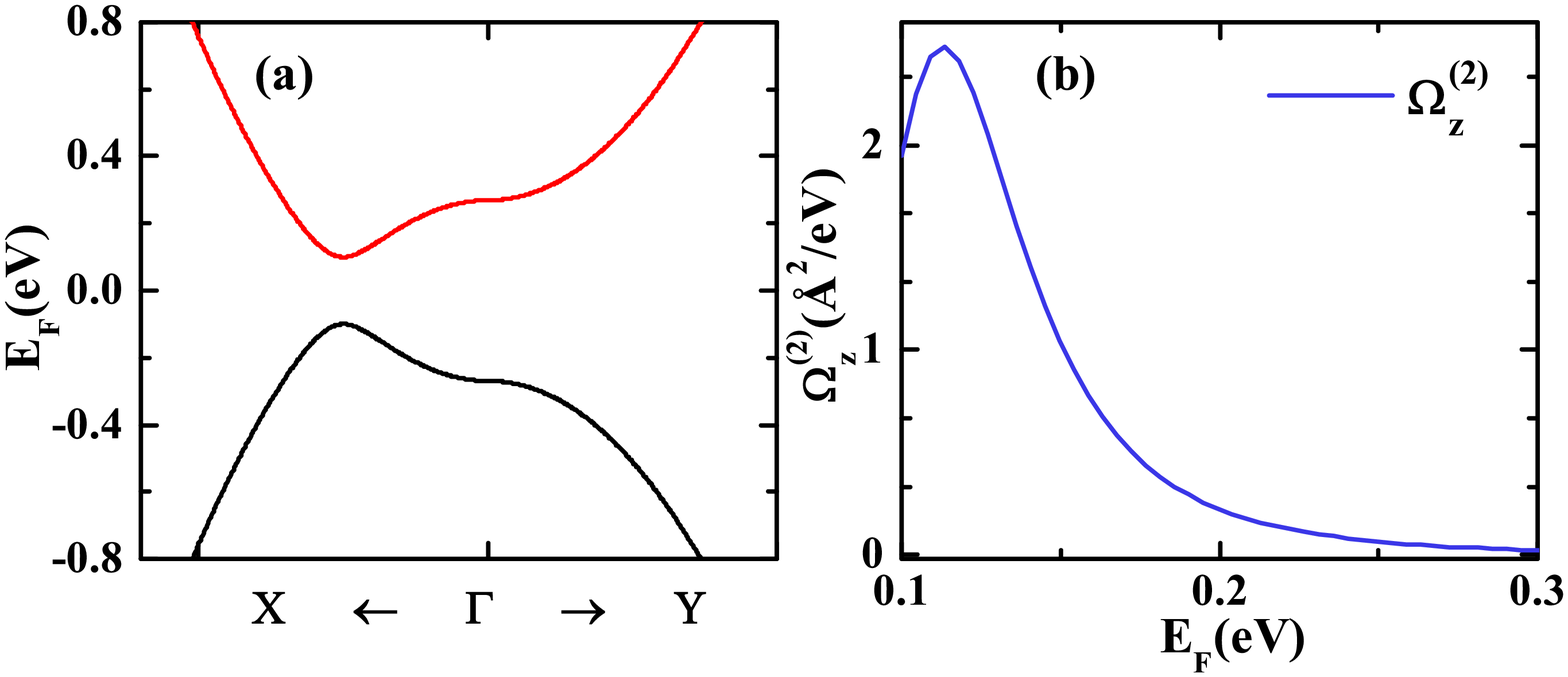}
\caption{(a) Band structure of the model system described by Eq.(\ref{ham}). (b) Berry curvature fluctuation $\Omega^{(2)}_z$ with respect to the Fermi energy. Parameters: $A=0$, $B=1$, $\delta=-0.25$, $v_2=1.0$, $d_0=0.1$\cite{L-Fu2}.}\label{fig1}
\end{figure}

In 2D systems with nonzero $\Omega_z$, only $\Omega^{(2)}_z=\int_k (-\partial_\epsilon f_0) \Omega_z^2$ exists as a positive scalar. In 3D, noncentrosymmetric point groups supporting nonvanishing $\Omega^{(2)}_{ab}$ are $\{C_1,C_{1v},C_{3h},C_n,C_{nv},D_{2d},D_{3h},D_n,S_4,T,T_d,O \}$ with $n=2,3,4,6$. The symmetry analysis is very similar to that in Ref. [\onlinecite{L-Fu}]. As an example, we show the energy dependence of $\Omega^{(2)}_z$ in Fig.\ref{fig1}(b) for a 2D model system, and discuss how to extract $\Omega^{(2)}_z$ from Hall noise signals later.

The third-order correlation $K_{abc}$ is explicitly expressed in Sec.{\color{blue}I}(3) of the supplementary material.\cite{sup} Following this route, full-counting statistics\cite{Levitov,Tang} within the Boltzmann approach is established. We also derive similar expressions of the linear and second-order Hall noises within the scattering matrix theory (SMT) to demonstrate the agreement between these two methods, as shown in Sec.{\color{blue}II} of the supplementary material\cite{sup} and references [\onlinecite{BG-Wang2,but2,Fisher}] therein. SMT is suitable for describing coherent nonlinear transport in multiterminal systems.\cite{WeiMM,WeiMM2}

We focus on the Hall noise spectrum, which can be simultaneously probed with the Hall currents in one measurement. Characteristics of the linear and second-order Hall noises will be discussed for the following model.

\bigskip
{\noindent{\bf \it 2D tilted massive Dirac model.}} The model Hamiltonian under investigation is,\cite{L-Fu2}
\begin{equation}
H\left( \mathbf{k} \right)=A{{k}^{2}}+\left( B{{k}^{2}}+\delta \right){{\sigma }_{z}}+{{v}_{2}}{{k}_{y}}{{\sigma }_{y}}+d_0{{\sigma }_{x}}, \label{ham}
\end{equation}
where $A$, $B$, $v_2$, $\delta$, and $d_0$ are system parameters, and ${{\sigma }_{x,y,z}}$ are Pauli matrices. This model breaks inversion symmetry $\cal I$ but preserves both $\cal T$ and mirror symmetry ${\cal M}_x$, hence BCD $D_x$ exists. Its band structure is shown in Fig.\ref{fig1}(a). Such a Hamiltonian describes 2D tilted massive Dirac systems\cite{L-Fu2,disorderNHE,tiltexpe} and captures low-energy band features of T$_d$-WTe$_2$ and topological crystalline insulator SnTe.\cite{L-Fu,L-Fu2}

\begin{figure}[tbp]
\centering
\includegraphics[width=1\columnwidth]{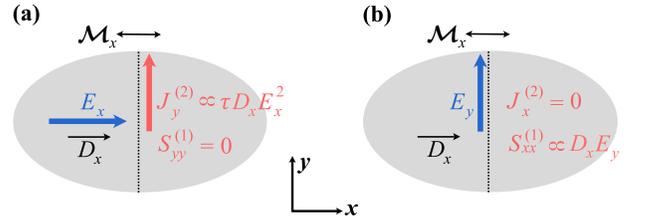}
\caption{Schematics of the Hall effects for the 2D tilted Dirac model. BCD $D_x$ (arrow vector) is orthogonal to the mirror line (dashed line). The electric field is applied along $x$ direction in (a) while along $y$ axis in (b). Intrinsic linear Hall noise $S^{(1)}_{H}$ and extrinsic second-order Hall current $J^{(2)}_H$ are expressed in Eq.(\ref{s1-m1}).}\label{fig2}
\end{figure}

For this model, the intrinsic linear Hall noises are
\begin{eqnarray}
S^{(1)}_{xx} &=& 2k_B T D_x E_y ~,~     J^{(2)}_x = 0; \nonumber \\
S^{(1)}_{yy} &=& 0~,~~~~~~~~~~~~~~~~~~~ J^{(2)}_y = -\tau D_x E_x^2. \label{s1-m1}
\end{eqnarray}
The second-order Hall currents are also shown for comparison. Here $S^{(1)}_{xx}$ is in response to the electric field in $y$ direction, while $J^{(2)}_{y}$ is driven by $E_x$, as demonstrated in Fig.\ref{fig2}. Band geometry $D_x$ can be extracted either from extrinsic Hall current $J^{(2)}_y$, or from intrinsic Hall noise $S^{(1)}_{xx}$. BCD-induced second-order Hall effect has been observed in large ranges of temperature and driving current for a variety of materials,\cite{NHEreview} and we expect BCD-induced linear Hall noise can be measured on similar platforms.\cite{S-Xu,NHEBi2Se3,H-Yang}

The phenomena displayed in Eq.(\ref{s1-m1}) appear to be counterintuitive: (1) when nonlinear Hall current $J^{(2)}_H$ vanishes, linear Hall noise is nonzero; (2) if $J^{(2)}_H$ is finite, linear noise $S^{(1)}_H$ is zero. It can be understood as follows: (1) $J^{(2)}_H = 0$ only means the average Hall current $\langle {\hat J}_H \rangle$ is zero up to the second order in electric field, but quantum correlation of currents is finite, i.e. $S^{(1)}_H \ne 0$; (2) when $J^{(2)}_H \ne 0$, only linear Hall noise $S^{(1)}_H$ vanishes,\cite{Fnote1} but the second-order Hall noise $S^{(2)}_H$ exists (as discussed below). Similar behaviors are reported in various nonequilibrium transport\cite{Altaner,Safi}: a system with pure spin current has finite charge noise when the average charge current vanishes\cite{BG-Wang1,Arakawa}; zero-current nonequilibrium delta-$T$ noise are generated by pure temperature bias.\cite{Lumbroso,Sivre,Larocque,Eriksson}

\begin{figure}[tbp]
\centering
\includegraphics[width=\columnwidth]{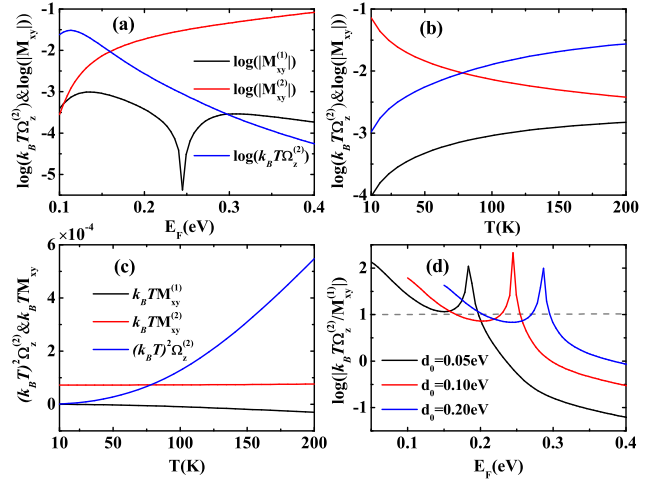}
\caption{The second-order Hall noise for the model system. Extrinsic noise $M_{xy}$ and intrinsic noise $k_B T\Omega^{(2)}_z$ as a function of the Fermi energy (a) or temperature (b). (c) Temperature scaled noises versus $T$. (d) $k_B T \Omega^{(2)}_z/|{M}^{(1)}_{xy}|$ with respect to $E_F$ for different $d_0$. Parameters are the same as Fig.\ref{fig1}. $\tau=1 ~ \rm fs$.\cite{fnote3} The unit of $M_{xy}$ is $\rm \AA^2 (eV)^2\tau^2/\hbar^2$, which equals to $2.2 ~ \rm \AA^2$ at $\tau = 1 ~ \rm fs$. $T=100 ~ \rm K$ in (a) and (d), while $E_F = 0.15 ~ \rm eV$ in (b) and (c). }\label{fig3}
\end{figure}

The second-order Hall noises for the model are
\begin{eqnarray}
S^{(2)}_{xx} &=& [{M}_{yx} + k_B T \Omega^{(2)}_z] E_y^2,  \\
S^{(2)}_{yy} &=& [{M}_{xy} + k_B T \Omega^{(2)}_z] E_x^2, \label{S2H} \\
{M}_{ab} &=& \tau^2\int_k [-k_B T \partial_\epsilon f_0 \partial^2_a v_b^2 + (\partial_\epsilon f_0)^2 v_a^2 v_b^2 ]. \label{s2-m1}
\end{eqnarray}
Without loss of generality, we set $E_x=E_y=1$ in the following discussion. There are two contributions in $S^{(2)}_{H}$. The first term ${M}_{ab}$ scales as $\tau^2$, hence we refer it as the {\it extrinsic Hall noise}, whose existence is also confirmed by SMT (Sec. {\color{blue}II}(2) of the supplementary material.\cite{sup}) The second intrinsic term is contributed by $\Omega^{(2)}_z$. It seems difficult to extract $\Omega^{(2)}_z$ from the second-order Hall noise due to the presence of ${M}_{ab}$. We propose two strategies to isolate $\Omega^{(2)}_z$. \newline
(1) Energy dependence. For $S^{(2)}_{yy}$ in Eq.(\ref{S2H}), we can split the extrinsic noise ${M}_{xy}$ into two parts: ${M}^{(1)}_{xy}=-\tau^2 k_B T \int_k \partial_\epsilon f_0 \partial^2_x v_y^2$ and ${M}^{(2)}_{xy}=\tau^2 \int_k (\partial_\epsilon f_0)^2 v_x^2 v_y^2$. Fig.\ref{fig3}(a) shows that both ${M}^{(1)}_{xy}$ and ${M}^{(2)}_{xy}$ are much smaller than $k_B T \Omega^{(2)}_{z}$ for small energies. Hence $\Omega^{(2)}_{z}$ is easily extracted in the small energy range. However, when ${M}^{(2)}_{xy}$ dominates the Hall noise for large energies, this strategy fails. \newline
(2) Temperature scaling. The temperature dependence of these noise terms are illustrated in Fig.\ref{fig3}(b), where ${M}^{(2)}_{xy}$ is inversely proportional to $T$ while both ${M}^{(1)}_{xy}$ and $k_B T \Omega^{(2)}_z$ are directly proportional to $T$. When we scale them by multiplying a factor $k_B T$, Fig.\ref{fig3}(c) shows $k_B T {M}^{(2)}_{xy}$ is largely a constant. At low temperature, e.g., $T=10~ \rm K$, contributions from ${M}^{(1)}_{xy}$ and $k_B T \Omega^{(2)}_z$ to $S^{(2)}_{yy}$ are negligible compared with ${M}^{(2)}_{xy}$. This motivates the following treatment: (a) measure the Hall noise ${S}^{(2)}_{yy}$ with respect to $T$; (b) ${M}^{(2)}_{xy}$ is subtracted from ${S}^{(2)}_{yy}$ since $k_B T {M}^{(2)}_{xy}$ is a constant obtained at $T=10 ~ \rm K$; (c) since $k_B T \Omega^{(2)}_z /|{M}^{(1)}_{xy}| \ge 10$\cite{Fnote2} in the temperature interval of Fig.\ref{fig3}(b), $\Omega^{(2)}_z$ is obtained:
\begin{eqnarray}
\Omega^{(2)}_z \approx \frac{1}{k_B T} ( \frac{ {S}^{(2)}_{yy} }{ E_x^2 } - {M}^{(2)}_{xy} ).
\end{eqnarray}
To ensure the applicability of this treatment, we analyze the competition between $k_B T \Omega^{(2)}_z$ and ${M}^{(1)}_{xy}$ with respect to $E_F$ for different $d_0$. $d_0$ tunes the Berry curvature of the model. In Fig.\ref{fig3}(d), with the increasing of $d_0$, energy windows fulfilling $k_B T \Omega^{(2)}_z /|{M}^{(1)}_{xy}| \ge 10$ always exist (above the dashed line). Therefore, this temperature scaling strategy for isolating $\Omega^{(2)}_z$ is widely applicable. Here we fix $\tau=1 ~ \rm fs$ to simplify the discussion. The temperature scaling strategy works even better if $\tau$ is temperature dependent. As shown in Sec.{\color{blue}IV}(3) of the supplementary material,\cite{sup} $\Omega^{(2)}_z$ is obtained through simple curve fitting when $\tau$ as a function of $T$ is established. For the model Hamiltonian of T$_d$-WTe$_2$, we numerically find $\Omega^{(2)}_z \sim 1 ~ \rm \AA^2/eV$ in Fig.\ref{fig1}(b). For SnTe, $\Omega^{(2)}_z \sim 0.01 ~ \rm \AA^2/eV$ (Sec.{\color{blue}IV}(1) of the supplementary material\cite{sup} and reference [\onlinecite{SnTeLattice}]).

\bigskip
\noindent{\it{Discussion.}} The thermal Hall noise spectrum and Berry curvature fluctuation can be investigated in experimental platforms which were previously adopted to study BCD-induced second-order Hall effect.\cite{S-Xu,Q-Ma,K-Kang,J-Xiao,NHEBi2Se3,H-Yang} The measurement of thermal noise has been developed as a standard technique for over two decades.\cite{Henny1999,Birk1995,Birk1996,crosscorrelation}

In 3D systems, Berry curvature may have more than one nonzero component and could induce an additional intrinsic term in the second-order Hall noise, which corresponds to cross-correlation of Berry curvature. We find
\begin{eqnarray}
S^{(2)}_{xx} &=& [{ M}_{yx}+k_B T \Omega^{(2)}_z] E_y^2+ [{ M}_{zx}+ k_B T \Omega^{(2)}_y] E_z^2 \nonumber \\
&+&  k_B T \Omega^{(2)}_{zy} E_y E_z. \label{cross}
\end{eqnarray}
With both $\cal T$ and ${\cal M}_x$ symmetries, $\Omega^{(2)}_{zy}=\int_k (-\partial_\epsilon f_0) \Omega_z \Omega_y $ is an even function of $k$ hence nonvanishes. To detect $S^{(2)}_{xx}$, the electric field is applied along ${\hat y}\cos\theta + {\hat z}\sin\theta$ direction. Rotating the electric field in $y$-$z$ plane by changing $\theta$, we can determine the three terms of $S^{(2)}_{xx}$. The first (second) term is probed at $\theta=0$ ($\theta=\pi/2$), and the third term is obtained by subtracting the other two terms from $S^{(2)}_{xx}$ measured at $\theta \ne 0,\pi/2$. Note that $\Omega^{(2)}_{zy}$ is directly accessible through this three-step measurement, and only observable in 3D.

In summary, by studying the thermal Hall noise spectrum of time-reversal invariant systems, we have found intrinsic Hall noises in both linear and nonlinear response regimes, which are all related to Berry curvature as well as quantum geometric tensor. The intrinsic linear Hall noise is proportional to Berry curvature dipole. The second-order Hall noise has intrinsic contribution from Berry curvature fluctuation. These intrinsic thermal Hall noises are the manifestation of quantum geometric tensor and its quantum fluctuation, which can be detected on platforms such as T$_d$-WTe$_2$.

We acknowledge support from the National Natural Science Foundation of China (Grants No. 12034014, No. 12004442, and No. 12174262).


\begin{thebibliography}{27}
\bibitem{Fnote0}
Since each band contributes independently to $S^{T}_{ab}$, we omit the band index and sum over all occupied bands. Meanwhile, the integration depends on the spatial dimensionality: $\int_k \equiv \int d^d k /(2\pi)^d$.
\bibitem{Fnote1}
The vanishing of $S_H^{(1)}$ can also be interpreted from the point view of symmetry, see Fig.S1 of the Supplimentary Material and related discussion.
\bibitem{Fnote2}
An error bar less than $10\%$ is allowed for this requirement, as long as there is major difference between these two terms.
\bibitem{fnote3}
In Fig.S17 of Ref. [\onlinecite{S-Xu}], the conductivity is around $\sigma=2 ~ \mu \rm S/m$ at $T \simeq 100 \rm K$ for monolayer WTe$_2$. From the Drude formula $\sigma= ne^2 \tau/{m^*}$ with $n = 10^{12} ~ \rm cm^{-3}$ and ${m^*}=0.3 ~ m_e$, the relaxation time $\tau$ is in the order of $1 ~ \rm fs$ for $T \leq 100 ~ \rm K$.
\end{thebibliography}
\end{document}